\documentclass{pasj00}

\newcounter{author}
\def\authorcount#1#2{\refstepcounter{author}\label{#1}
                     \altaffiltext{\ref{#1}}{#2}}

\begin{document}
\SetRunningHead{T. Kato et al.}{AM CVn-Type Object SDSS J090221.35$+$381941.9}

\Received{201X/XX/XX}
\Accepted{201X/XX/XX}

\title{Superoutburst of SDSS J090221.35$+$381941.9:
       First Measurement of Mass Ratio in an AM CVn-Type Object
       using Growing Superhumps}

\author{Taichi~\textsc{Kato},\altaffilmark{\ref{affil:Kyoto}*}
        Tomohito~\textsc{Ohshima},\altaffilmark{\ref{affil:Kyoto}}
        Denis~\textsc{Denisenko},\altaffilmark{\ref{affil:Denisenko}}
        Pavol~A.~\textsc{Dubovsky},\altaffilmark{\ref{affil:Dubovsky}}
        Igor~\textsc{Kudzej},\altaffilmark{\ref{affil:Dubovsky}}
        William~\textsc{Stein},\altaffilmark{\ref{affil:Stein}}
        Enrique~de~\textsc{Miguel},\altaffilmark{\ref{affil:Miguel}}$^,$\altaffilmark{\ref{affil:Miguel2}}
        Arne~\textsc{Henden},\altaffilmark{\ref{affil:AAVSO}}
        Ian~\textsc{Miller},\altaffilmark{\ref{affil:Miller}}
        Kirill~\textsc{Antonyuk},\altaffilmark{\ref{affil:CrAO}}
        Oksana~\textsc{Antonyuk},\altaffilmark{\ref{affil:CrAO}}
        Nikolaj~\textsc{Pit},\altaffilmark{\ref{affil:CrAO}}
        Aleksei~\textsc{Sosnovskij},\altaffilmark{\ref{affil:CrAO}}
        Alex~\textsc{Baklanov},\altaffilmark{\ref{affil:CrAO}}
        Julia~\textsc{Babina},\altaffilmark{\ref{affil:CrAO}}
        Elena~P.~\textsc{Pavlenko},\altaffilmark{\ref{affil:CrAO}}$^,$\altaffilmark{\ref{affil:Kyoto}}
        Kazunari~\textsc{Matsumoto},\altaffilmark{\ref{affil:OKU}}
        Daiki~\textsc{Fukushima},\altaffilmark{\ref{affil:OKU}}
        Megumi~\textsc{Takenaka},\altaffilmark{\ref{affil:OKU}}
        Miho~\textsc{Kawabata},\altaffilmark{\ref{affil:OKU}}
        Daisuke~\textsc{Daisuke},\altaffilmark{\ref{affil:OKU}}
        Kazuki~\textsc{Maeda},\altaffilmark{\ref{affil:OKU}}
        Risa~\textsc{Matsuda},\altaffilmark{\ref{affil:OKU}}
        Katsura~\textsc{Matsumoto},\altaffilmark{\ref{affil:OKU}}
        Colin~\textsc{Littlefield},\altaffilmark{\ref{affil:LCO}}
        Arto~\textsc{Oksanen},\altaffilmark{\ref{affil:Nyrola}}
        Hiroshi~\textsc{Itoh},\altaffilmark{\ref{affil:Ioh}}
        Gianluca~\textsc{Masi},\altaffilmark{\ref{affil:Masi}}
        Francesca~\textsc{Nocentini},\altaffilmark{\ref{affil:Masi}}
        Patrick~\textsc{Schmeer},\altaffilmark{\ref{affil:Schmeer}}
        Roger~D.~\textsc{Pickard},\altaffilmark{\ref{affil:BAAVSS}}$^,$\altaffilmark{\ref{affil:Pickard}}
        Seiichiro~\textsc{Kiyota},\altaffilmark{\ref{affil:Kis}}
        Shawn~\textsc{Dvorak},\altaffilmark{\ref{affil:Dvorak}}
        Joseph~\textsc{Ulowetz},\altaffilmark{\ref{affil:Ulowetz}}
        Yutaka~\textsc{Maeda},\altaffilmark{\ref{affil:Mdy}}
        Ra\'ul~\textsc{Michel},\altaffilmark{\ref{affil:UNAM}}
        Sergey~Yu.~\textsc{Shugarov},\altaffilmark{\ref{affil:Sternberg}}$^,$\altaffilmark{\ref{affil:Slovak}}
        Drahomir~\textsc{Chochol},\altaffilmark{\ref{affil:Slovak}}
        Rudolf~\textsc{Nov\'ak},\altaffilmark{\ref{affil:Novak}}
}

\authorcount{affil:Kyoto}{
     Department of Astronomy, Kyoto University, Kyoto 606-8502}
\email{$^*$tkato@kusastro.kyoto-u.ac.jp}

\authorcount{affil:Denisenko}{
     Space Research Institute (IKI), Russian Academy of Sciences, Moscow,
     Russia}

\authorcount{affil:Dubovsky}{
     Vihorlat Observatory, Mierova 4, Humenne, Slovakia}

\authorcount{affil:Stein}{
     6025 Calle Paraiso, Las Cruces, New Mexico 88012, USA}

\authorcount{affil:Miguel}{
     Departamento de F\'isica Aplicada, Facultad de Ciencias
     Experimentales, Universidad de Huelva,
     21071 Huelva, Spain}

\authorcount{affil:Miguel2}{
     Center for Backyard Astrophysics, Observatorio del CIECEM,
     Parque Dunar, Matalasca\~nas, 21760 Almonte, Huelva, Spain}

\authorcount{affil:AAVSO}{
     American Association of Variable Star Observers, 49 Bay State Rd.,
     Cambridge, MA 02138, USA}

\authorcount{affil:Miller}{
     Furzehill House, Ilston, Swansea, SA2 7LE, UK}

\authorcount{affil:CrAO}{
     Crimean Astrophysical Observatory, Kyiv Shevchenko 
     National University, 98409, Nauchny, Crimea, Ukraine}

\authorcount{affil:OKU}{
     Osaka Kyoiku University, 4-698-1 Asahigaoka, Osaka 582-8582}

\authorcount{affil:LCO}{
     Department of Physics, University of Notre Dame, Notre Dame,
     Indiana 46556, USA}

\authorcount{affil:Nyrola}{
     Hankasalmi observatory, Jyvaskylan Sirius ry, Vertaalantie
     419, FI-40270 Palokka, Finland}

\authorcount{affil:Ioh}{
     Variable Star Observers League in Japan (VSOLJ),
     1001-105 Nishiterakata, Hachioji, Tokyo 192-0153}

\authorcount{affil:Masi}{
     The Virtual Telescope Project, Via Madonna del Loco 47, 03023
     Ceccano (FR), Italy}

\authorcount{affil:Schmeer}{
     Bischmisheim, Am Probstbaum 10, 66132 Saarbr\"{u}cken, Germany}

\authorcount{affil:BAAVSS}{
     The British Astronomical Association, Variable Star Section (BAA VSS),
     Burlington House, Piccadilly, London, W1J 0DU, UK}

\authorcount{affil:Pickard}{
     3 The Birches, Shobdon, Leominster, Herefordshire, HR6 9NG, UK}

\authorcount{affil:Kis}{
     VSOLJ, 7-1 Kitahatsutomi, Kamagaya, Chiba 273-0126}

\authorcount{affil:Dvorak}{
     Rolling Hills Observatory, 1643 Nightfall Drive,
     Clermont, Florida 34711, USA}

\authorcount{affil:Ulowetz}{
     Center for Backyard Astrophysics Illinois,
     Northbrook Meadow Observatory, 855 Fair Ln, Northbrook,
     Illinois 60062, USA}

\authorcount{affil:Mdy}{
     Kaminishiyamamachi 12-14, Nagasaki, Nagasaki 850-0006}

\authorcount{affil:UNAM}{
     Instituto de Astronom\'{\i}a UNAM, Apartado Postal 877, 22800 Ensenada
     B.C., M\'{e}xico}

\authorcount{affil:Sternberg}{
     Sternberg Astronomical Institute, Lomonosov Moscow University, 
     Universitetsky Ave., 13, Moscow 119992, Russia}

\authorcount{affil:Slovak}{
     Astronomical Institute of the Slovak Academy of Sciences, 05960,
     Tatranska Lomnica, the Slovak Republic}

\authorcount{affil:Novak}{
     Research Centre for Toxic Compounds in the Environment, Faculty of 
     Science, Masaryk University, Kamenice 3, 625 00 Brno, Czech Republic}



\maketitle

\begin{abstract}
   We report on a superoutburst of the AM CVn-type object
SDSS J090221.35$+$381941.9 [J0902; orbital period 0.03355(6) d]
in 2014 March--April.
The entire outburst consisted of a precursor outburst
and the main superoutburst, followed by a short rebrightening.
During the rising branch of the main superoutburst, we detected
growing superhumps (stage A superhumps) with a period of
0.03409(1)~d.  During the plateau phase of the superoutburst,
superhumps with a shorter period (stage B superhumps) were
observed.  Using the orbital period and the period of the
stage A superhumps, we were able to measure the dynamical
precession rate of the accretion disk at the 3:1 resonance,
and obtained a mass ratio ($q$) of 0.041(7).  This is the first
successful measurement of the mass ratio in an AM CVn-type
object using the recently developed stage A superhump method.
The value is generally in good agreement with the theoretical
evolutionary model.  The orbital period
of J0902 is the longest among the outbursting AM CVn-type
objects, and the borderline between the outbursting systems
and systems with stable cool disks appears to be longer
than had been supposed.
\end{abstract}

\section{Introduction}

   AM CVn-type objects are a class of cataclysmic variables (CVs)
composed of an accreting white dwarf and a mass-transferring
helium white dwarf (secondary star).  The orbital periods
($P_{\rm orb}$) of AM CVn-type objects are in a range 
of 5 and 65 min, and comprise the population of the
shortest-$P_{\rm orb}$ CVs [for recent reviews of
AM CVn-type objects, see e.g. \citet{nel05amcvnreview};
\citet{sol10amcvnreview}].

   AM CVn-type objects have recently been becoming an interesting
topic in astrophysics because some of them are considered
as the most promising objects for direct detection of
the gravitational wave radiation (e.g. \cite{nel03amcvnGWR}).
Some AM CVn-type objects are also considered as progenitors
of a population of type-Ia supernovae \citep{sol05amcvnSN}.
Three evolutionary paths have been proposed to form
AM CVn-type objects, and they have been widely discussed
both in terms of theoretical population synthesis and
observations (see e.g. \cite{sol10amcvnreview}).

   Still, basic parameters of
AM CVn-type objects, such as mass ratios ($q=M_2/M_1$),
are difficult to measure observationally and this difficulty
has hindered the comparison between the theory and observation.
This difficulty partly comes from the difficulty in directly
detecting the secondary and there have been attempts to
estimate the mass ratios by using the Doppler tomography
of the accretion disk (\cite{mar99gpcom}; \cite{nel01amcvnspot}; 
\cite{roe06amcvn}).

   Many AM CVn-type objects show superhumps, which
arise as a result of the precession of the eccentric accretion
disk deformed by the 3:1 resonance with the secondary 
(\cite{whi88tidal}; \cite{hir90SHexcess}). The fractional superhump
excess ($\varepsilon \equiv P_{\rm SH}/P_{\rm orb}-1$,
where $P_{\rm SH}$ is the superhump period) reflects the precession rate,
and the empirical relation between $\varepsilon$ and $q$
(such as \cite{pat98evolution}, \cite{pat05SH}), which has been
developed and calibrated for hydrogen-rich CVs, has been used to 
estimate the $q$ values.

   This method, however, suffers from the unknown degree
of the pressure effect (e.g. \cite{pea07amcvnSH}).
There has been only one direct measurement of the $q$ value
in an AM CVn-type object by analyzing eclipses in
SDSS J092620.42$+$034542.3 \citep{cop11j0926}.
The only other known eclipsing AM CVn-type object,
PTF1 J191905.19$+$481506.2, is only partially eclipsing
and the $q$ value has not been determined \citep{lev14j1919}.

   In the last two years, however, there has been a great
progress in understanding the precession rate in the
superhumping accretion disk, and it has become clear
that the superhumps in the growing stage during
the superoutburst reflects the dynamical precession rate
at the radius of the 3:1 resonance \citep{kat13qfromstageA}.
Although this method was initially developed for hydrogen-rich
CVs, it is expected to be applicable to AM CVn-type objects
since it only depends on dynamics.
Here, we present the first successful result.

\section{SDSS J090221.35$+$381941.9}

   SDSS J090221.35$+$381941.9 (hereafter J0902) is an AM CVn-type
object selected by color using the Sloan Digital Sky Survey (SDSS)
\citep{rau10HeDN}.  \citet{rau10HeDN} obtained time-resolved
spectra of this object and identified the orbital period
of 48.31(8) min [0.03355(6)~d] from the radial velocity variations of
the emission lines.  Although the continuum could be well reproduced
by a blackbody of 15000~K, the lack of the broad absorption
lines suggested that the accreting white dwarf is either cooler
than 15000~K or the additional component contributes to
the continuum.

   The emission-line spectrum in \citet{rau10HeDN}
was indicative of an AM CVn-type object with a cold, quiescent disk,
unlike a thermally stable disk as in AM CVn.
No outburst was recorded in this system in the past.  

   On 2014 March 6, D. Denisenko detected an outburst from
the MASTER-Kislovodsk (see \cite{MASTER} for the MASTER
network) images (vsnet-alert 16982).\footnote{
VSNET-alert archive can be accessed at
$<$http://ooruri.kusastro.kyoto-u.ac.jp/pipermail/vsnet-alert/$>$.
}
The object experienced a very rapid fading (vsnet-alert 16986,
16988) at a rate of 2.3--2.9 mag d$^{-1}$.
This outburst turned out to be a precursor outburst.
The object started rising on March 12 (vsnet-alert 17016)
and went into a superoutburst on March 15--16 accompanied 
by developing superhumps (vsnet-alert 17023).

\section{Observation and Analysis}\label{sec:obs}

   The data were acquired by time-resolved unfiltered and $V$-band
CCD photometry using 30--40cm class telescopes 
by the VSNET Collaboration \citep{VSNET} and
the public data from the AAVSO International Database.\footnote{
   $<$http://www.aavso.org/data-download$>$.
}.
All the observed times were corrected to Barycentric Julian Date (BJD).
Before making the analysis, we corrected zero-point differences
between different observers by adding a constant to each observer.
For the outbursting CVs, the magnitude system of the unfiltered
CCD observations is close to $V$.
The details of the observations will be presented in a separate paper.

   The data analysis was performed just in the same way described
in \citet{Pdot} and \citet{Pdot3}.  
The times of superhump maxima were determined using the template fitting 
method as described in \citet{Pdot} after de-trending
the global variation due to the outburst by using
the locally-weighted polynomial regression (LOWESS: \cite{LOWESS}).
The superhump periods were determined using 
phase dispersion minimization (PDM; \cite{PDM})
for period analysis and 1$\sigma$ errors for the PDM analysis
were estimated by the methods of \citet{fer89error} and \citet{Pdot2}.

\section{Results}\label{sec:result}

\subsection{Outburst}\label{sec:out}

\begin{figure}
  \begin{center}
    \FigureFile(88mm,70mm){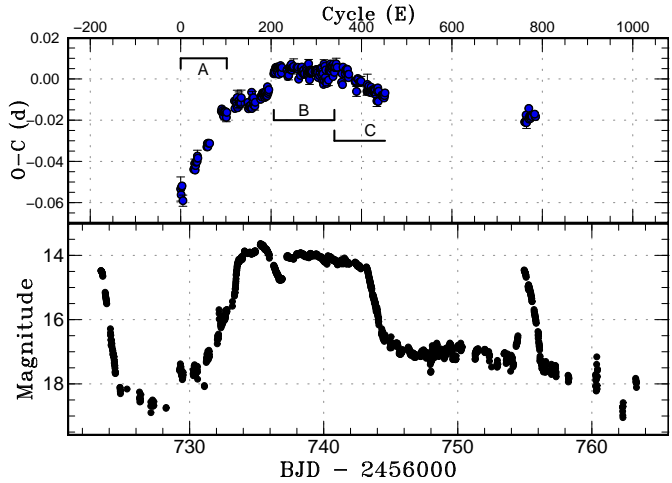}
  \end{center}
  \caption{$O-C$ diagram and light curve of J0902.
  (Upper) $O-C$ diagram.  An ephemeris of Max(BJD)$=2456729.3312+0.03372E$
  was used to draw this figure.  The intervals A--C represents
  superhump stages (see text for detail).  The cycle counts
  between the main superoutburst and the rebrightening
  are uncertain.
  (Lower) Light curve.  The observations were binned to
  0.01~d.
  }
  \label{fig:j0902humpall}
\end{figure}

   Following the precursor outburst, the object remained
in a state $\sim$2 mag brighter than quiescence for 4~d
(lower panel of figure \ref{fig:j0902humpall}).
The object then started to brighten slowly, and the final
rise to a superoutburst took place on BJD 2456733.
After reaching the temporary maximum, the object again
started fading (BJD 2456735--2456737).  This fading episode
(we call it a ``dip'') was,
however, temporary and the object then entered a plateau
phase of the superoutburst for the next 6~d.
The object then started fading from the superoutburst
on BJD 2456743 at a rate of $\sim$1.8 mag d$^{-1}$.
The object then remained around magnitude 17 ($\sim$3 mag
brighter than quiescence) for 9~d.
There was at least one post-superoutburst rebrightening
on BJD 2456754--2456755.  The fading from this
rebrightening was also very rapid ($\sim$2.1 mag d$^{-1}$).

\subsection{Superhumps}\label{sec:sh}

\begin{figure}
  \begin{center}
    \FigureFile(88mm,90mm){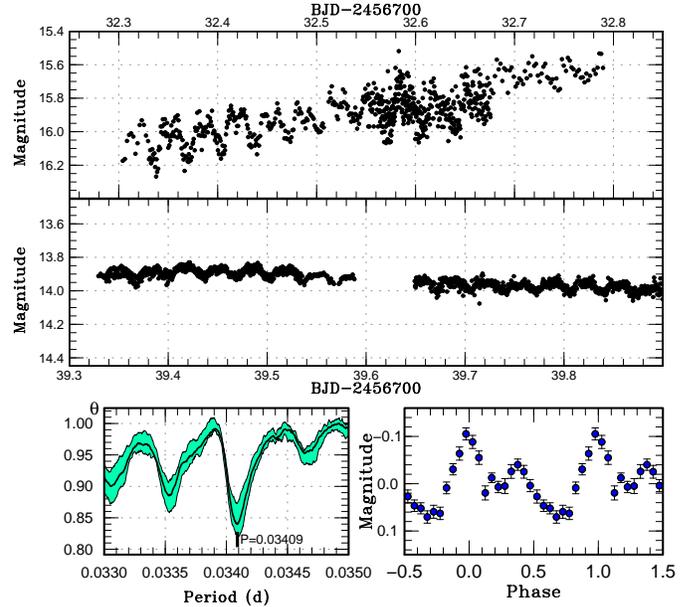}
  \end{center}
  \caption{Superhumps in J0902.
  (Upper) Superhumps in the rising stage of the superoutburst
  (stage A superhumps).
  (Middle) Superhumps after the ``dip'' phenomenon early in the main
  superoutburst (stage B superhumps).
  (Lower left) PDM analysis of stage A superhumps.
  (Lower right) Mean profile of stage A superhumps. 
  }
  \label{fig:j0902humpgrow}
\end{figure}

   Superhumps started to appear during the slowly rising
phase following the precursor (upper panel of figure
\ref{fig:j0902humpgrow}).  The superhumps observed
in this interval had a long period, and this long period
was observed for $\sim$102 cycles (upper panel of figure
\ref{fig:j0902humpall}).  During the maximum before
the temporary dip in the early part of the superoutburst,
the period variation became more complex and the orbital period
was detected as a transient signal.  Following this
temporary dip, the superhump period stabilized to
a shorter period (lower panel of figure
\ref{fig:j0902humpgrow}).  There was a jump in the phase
following the dip.  The phase jump was significantly
smaller than 0.5 phase, and this jump did not resemble
a transition to the so-called ``traditional''
late superhumps \citep{vog83lateSH}.
After a further $\sim$140 cycles, the period again
decreased discontinuously.

   Based on the similarity of the $O-C$ diagram
to those of recently identified candidate period bouncers
in hydrogen-rich CVs (\cite{kat13j1222}; \cite{nak14j0754j2304}),
we identified $E=0$--102 (see upper panel of figure
\ref{fig:j0902humpall}) as the growing stage of superhumps
(stage A) and $206 \le E \le 340$ as stage B and
$E \ge 340$ as stage C (for the superhump stages and
the typical behavior in hydrogen-rich systems, see \cite{Pdot}).
Although superhumps were
continuously seen after the fading from the superoutburst
plateau, the individual times of maxima are not
shown on this figure due to the large errors 
in determination.  There was a transition phase
between stage A and B, which is unique to this object.
The mean superhump periods in stages A, B and C were
0.03409(1)~d, 0.03371(1)~d and 0.03359(1)~d,
respectively.

\section{Discussion}\label{sec:discussion}

\subsection{Outburst in a Long-Period AM Canum Venaticorum Star}
\label{sec:outburst}

   Mass transfer in AM CVn-type objects is believed 
to be mainly powered by the gravitational wave radiation.
Since the secondary is degenerate, the orbital period becomes longer
as the secondary transfers the matter.  As the systems evolve,
the mass-transfer rates quickly decrease (\cite{tsu97amcvn};
\cite{nel01amcvnpopulationsynthesis}).  It is known that helium
accretion disks are prone to thermal instability, as well as in
hydrogen-rich CVs, and systems with intermediate mass-transfer rates
show dwarf nova (DN) type outbursts \citep{tsu97amcvn}.
This condition is usually considered to be achieved in
systems with orbital periods of 20--40 min \citep{nel05amcvnreview}.
Long-$P_{\rm orb}$ objects like GP Com ($P_{\rm orb}$=46.57 min)
have never shown an outburst and are considered to have sufficiently
low mass-transfer rates and possess stable cool disks.
These expectations have generally been confirmed by observations
(e.g. \cite{ram12amcvnLC}).

   J0902 has the longest $P_{\rm orb}$ among AM CVn-type
objects which have ever shown outbursts [the previous record
was CSS J045019.7$-$093113 having a superhump period of 47.28(1) min
with some uncertainty in alias selection \citep{wou13j0450atel4726}].
\citet{nel05amcvnreview} suggested that the transition to 
the stable cool disks
(without dwarf nova outbursts) happens between orbital periods
of 34 and 39 min.  The existence of two dwarf nova-type
objects (J0902 and CSS J045019.7$-$093113) indicates that
this transition happens in longer orbital periods (47--48 min).
It would be an interesting question whether GP Com may undergo 
an outburst or whether GP Com has different properties
despite the similarity of its orbital period with J0902.

\subsection{Slow Evolution of Superhumps}\label{sec:shevolution}

   In this object, it took more than 100 cycles
to develop fully grown superhumps.
AM CVn-type systems (at least for object with long
orbital periods) are expected to have a very light
secondary, and the small mass ratio is most likely
responsible for the long duration of stage A
since the growth rate of the superhumps is known
to be proportional to $q^2$ (\cite{lub91SHa}; see a discussion in
\cite{kat13j1222}).  This finding is also compatible
with the $q$ estimation in the following subsection.

\subsection{Estimation of the Mass Ratio from Stage A Superhumps}
\label{sec:qestimate}

   According to \citet{kat13qfromstageA}, the precession
frequency of stage A superhumps reflects the dynamical
precession rate of the eccentric disk at the radius
of the 3:1 resonance.  The fractional superhump excess
in frequency unit $\varepsilon^* \equiv 1-P_{\rm orb}/P_{\rm SH}$
for stage A superhumps is 0.0158(18).  This value
corresponds to $q$=0.041(7) (see table 1 or figure 2 in
\cite{kat13qfromstageA}).  This result became the first
measurement of the $q$ value by this method in an AM CVn-type system.
Since most of the error in $q$ comes from the uncertainty
in $P_{\rm orb}$, this $q$ value will be improved
by further refinement of $P_{\rm orb}$.

\begin{figure}
  \begin{center}
    \FigureFile(88mm,70mm){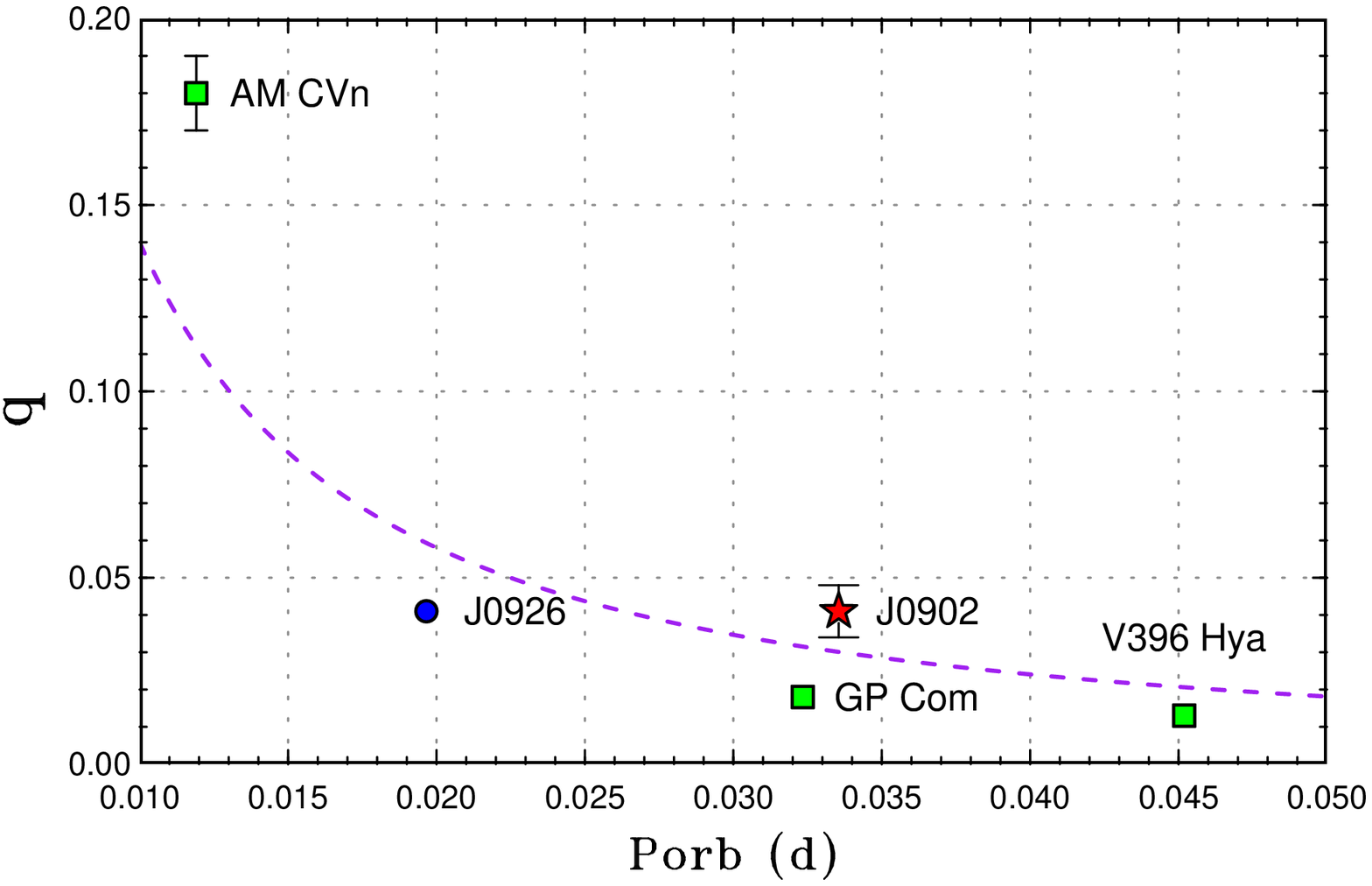}
  \end{center}
  \caption{Comparison of mass ratios in AM CVn-type objects.
  The filled circle represents the measurement from eclipse
  observations.  The filled squares represent the measurements
  from Doppler tomography (AM CVn: \cite{roe06amcvn};
  GP Com: \cite{mar99gpcom};
  V396 Hya: Steeghs et al., in preparation, from \cite{sol10amcvnreview}).
  The location of J0902 is shown with a filled star.
  The dashed curve represents equation (9) in \citet{tsu97amcvn}
  and assumed a 0.8 $M_\odot$ primary.
  }
  \label{fig:evolseq}
\end{figure}

   The $q$ value of 0.041(2) for the eclipsing
system SDSS J092620.42$+$034542.3 
($P_{\rm orb}$=28.31 min) is comfortably close to 
the current estimate of J0902 (figure \ref{fig:evolseq}).
The location of J0902 seems to be consistent with
the theoretical evolutionary track representing the mass-radius
relation for a semi-degenerate secondary [\citet{tsu97amcvn};
see e.g. \citet{yun08HeCV} for more detailed modeling].

   The commonly used relation between
fractional superhump excess and $q$ [such as
\citet{pat98evolution}, \citet{pat05SH}, \citet{Pdot}]
has uncertain errors due to the unknown pressure effect
(cf. \cite{kat13qfromstageA}), which is
particularly the case for AM CVn-type objects
\citep{pea07amcvnSH}.  The value of $\varepsilon^*$=0.0047(18) for 
stage B superhumps in J0902 corresponds to $q$=0.025(10)
by the traditional method in \citet{pat05SH}; it is
probable that the traditional method highly underestimate
$q$ values in AM CVn-type objects.  We therefore did not
include $q$ values from superhumps (other than stage A)
in figure \ref{fig:evolseq}.
We propose that stage A superhumps
provide an efficient and reliable tool for studying 
the evolutionary track of AM CVn-type objects.

\subsection{Pressure Effect in Helium Disks}
\label{sec:pressureeffect}

   \citet{pea07amcvnSH} suggested that the pressure effect
in helium disks may be higher than in hydrogen-rich
disk since the ionization temperature is higher.

   According to \citet{lub92SH}, the apsidal precession frequency
($\nu_{\rm pr}$) can be written as a form:
\begin{equation}
\nu_{\rm pr}=\nu_{\rm dyn}+\nu_{\rm pressure}+\nu_{\rm stress},
\label{equ:Lubows}
\end{equation}
where the first term, $\nu_{\rm dyn}$, represents a contribution 
to disk precession due to the dynamical force of the secondary, 
giving rise to prograde precession; the second term, $\nu_{\rm pressure}$, 
the pressure effect giving rise to retrograde precession; and 
the last term, $\nu_{\rm stress}$, the minor wave-wave interaction.
If we neglect the last term, we can estimate the contribution
of the pressure effect by estimating
$\nu_{\rm pressure}/\nu_{\rm orb}$, where $\nu_{\rm orb}$
is the orbital frequency.  In stage A, only $\nu_{\rm dyn}$
contributes to the disk precession and in stage B,
the combined effect of $\nu_{\rm dyn}$ and $\nu_{\rm pressure}$
is expected.  We can thus estimate $\nu_{\rm pressure}/\nu_{\rm orb}$
by evaluating the difference between $\varepsilon^*$ for
stages A and B (see also \cite{nak14j0754j2304}).

   In J0902, $\varepsilon^*$(stage A)$-\varepsilon^*$(stage B)
is 0.012, which is not particularly larger than
those in hydrogen-rich systems [0.010--0.015, figure 16
in \citet{Pdot}; see also \citet{nak13j2112j2037}].
It will be worth pointing out that the superhump period
for stage B superhumps may be shorter than the orbital period
if $q$ is sufficiently small.  This is expected to happen
for $q \le 0.03$ \citep{kat13qfromstageA}.  If AM CVn-type
dwarf novae with longer $P_{\rm orb}$ were to undergo
superoutbursts, such a situation may be observed.
The contribution of the pressure effect in helium disks
needs to be examined further using more samples.

\subsection{Transient Appearance of Orbital Signal}
\label{sec:orbitalsignal}

   During the rising branch to the superoutburst maximum,
a signal with a period very close to the orbital period
was detected (vsnet-alert 17043).  This signal was at the time
considered to be early superhumps, which are supposed to
arise from the 2:1 resonance in very low-$q$ systems
(cf. \cite{osa02wzsgehump}; \cite{kat02wzsgeESH}).
A similar period was also recorded during the late
stage of the plateau phase and a competition between
the 3:1 and 2:1 resonance was suggested (vsnet-alert 17082).

\begin{figure}
  \begin{center}
    \FigureFile(88mm,70mm){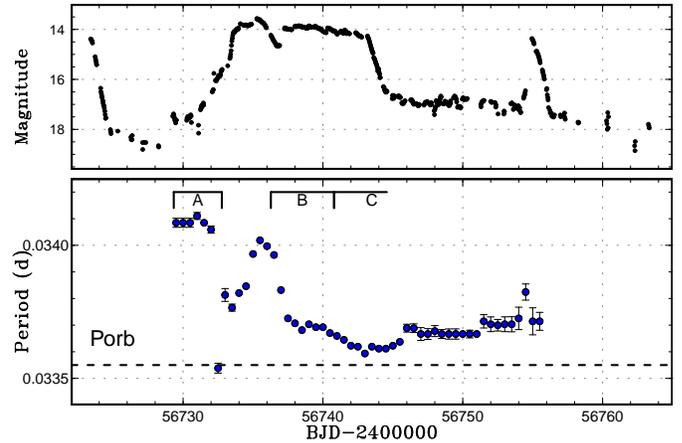}
  \end{center}
  \caption{Variation of the superhump period of J0902.
  (Upper) Light curve.  The observations were binned to
  0.03~d.
  (Lower) Superhump periods.  The periods were determined
  from 3-d segments (shifted by 0.5 d) using the PDM method.
  The initial long period corresponds to stage A superhumps.
  There was a phase of transition between stage A and stage B,
  during which the period of superhumps showed a large
  variation.  Superhump stages A--C are the same as in
  figure \ref{fig:j0902humpall}.  The start of stage B and
  the distinction between stage B and C are not as clearly
  defined as in the $O-C$ diagram (figure \ref{fig:j0902humpall})
  because the variation of the period is smeared by the use
  of 3-d segments.
  }
  \label{fig:j0902shvar}
\end{figure}

   We re-examined this behavior.  The variation of
the superhump periods is shown in figure \ref{fig:j0902shvar}.
The initial long period corresponds to stage A superhumps.
There was a phase of transition between stage A and stage B,
during which the period of superhumps showed a large
variation.  The initial appearance of the orbital signal
corresponds to this phase (BJD 2456732).  The second
appearance was on a smooth continuation of stage B to C
superhumps (around BJD 2456743).  It is evident from
this figure that the reported second appearance of 
the orbital signal was on the smooth extension of
ordinary superhumps, and the period was slightly different from
the orbital one.  A combination of a very small $q$,
shrinkage of the disk radius and the pressure effect
apparently reduced the precession rate very close
to zero.  The situation for the initial appearance
is less clear.  Since the period was so close to the
orbital period and the subsequent variation of 
the periods and the light curve were not regular, 
it may be indeed the case that the competition with 
the 2:1 resonance played some role in this phase.
The profile of the superhumps in this phase, however,
was not doubly humped as are commonly seen in
early superhumps \citep{kat02wzsgeESH}.

   The lack of long-lasting phase of early superhumps,
which are commonly seen in WZ Sge-type dwarf novae
(low-$q$, hydrogen-rich dwarf novae), can be understood
considering that the main superoutburst apparently
started as an inside-out-type (slowly rising)
outburst, and the disk could not expand sufficiently
to fully establish the 2:1 resonance.  Such a phenomenon
may be similar to those in candidate period bouncers, 
which are very low-$q$ hydrogen-rich systems 
\citep{nak14j0754j2304}.

\medskip

This work was supported by the Grant-in-Aid
``Initiative for High-Dimensional Data-Driven Science through Deepening
of Sparse Modeling'' from the Ministry of Education, Culture, Sports, 
Science and Technology (MEXT) of Japan.
We acknowledge with thanks the variable star
observations from the AAVSO International Database contributed by
observers worldwide and used in this research.
K. Antonyuk and N. Pit express a specific acknowledgement to
the funding of the CCD Camera FLI ProLine PL230 by Labex OSUG@2020.

\end{document}